\documentstyle[aps,prd,twocolumn]{revtex}
\begin{document}

\input epsfig.sty

\title{Casimir energy in a small volume multiply connected static 
hyperbolic pre-inflationary Universe}
\author{ Daniel M\"uller\thanks{Electronic Address: 
muller@andromeda.iagusp.usp.br}}
\address{{\small {\it Instituto Astron\^omico e Geof\'\i sico - USP}}\\
{\small  {\it Av. Miguel St\'efano, 4200, CEP-04301-904, 
S\~ao Paulo, SP, Brazil}}}
\author{Helio V. Fagundes\thanks{Electronic Address: helio@ift.unesp.br}}
\address{{\small {\it Instituto de F\'\i sica Te\'orica - UNESP}}\\
{\small {\it R. Pamplona, 145, CEP 01405-000, S\~ao Paulo, SP, Brazil}}}
\author{Reuven Opher\thanks{Electronic Address: opher@orion.iagusp.usp.br}}
\address{{\small {\it Instituto Astron\^omico e Geof\'\i sico - USP}}\\
{\small  {\it Av. Miguel St\'efano, 4200, CEP-04301-904, 
S\~ao Paulo, SP, Brazil}}}

\maketitle
\begin{abstract}
A few years ago, Cornish, Spergel and Starkman (CSS), 
suggested that a multiply connected ``small'' Universe could allow for 
classical chaotic mixing as a pre-inflationary homogenization process.  
The smaller the volume, the more important the process. Also, a smaller Universe 
has a greater probability of being spontaneously created. Previously DeWitt, 
Hart and Isham (DHI) calculated the Casimir energy for static 
multiply connected flat space-times. Due to the interest in small volume 
hyperbolic Universes (e.g. CSS), we generalize the DHI calculation by making a
a numerical investigation of the Casimir energy for a 
conformally coupled, massive scalar field in a static Universe, whose spatial 
sections are the Weeks manifold, the smallest Universe of negative curvature 
known. In spite of being a numerical calculation, our result is in fact exact. 
It is shown that there is spontaneous vacuum excitation of low multipolar components.

$\,\,$

\centerline{PACS number: 98:80.Cq, 98.80Hw}
\end{abstract}
\section{Introduction}

There has been an increase of interest in the topology of the Universe in
recent years. As is well known, the Einstein equations (EQ) restrict the
spatial homogeneous and isotropic sections to either $R^3,S^3$ or 
$H^3$, locally. 
Recent observational data indicate that the curvature of the Universe is 
small, without ruling out the negatively curved case. On the other
hand, the EQ are insensitive to the global nontrivial topology, induced by a
discrete group of isometries $\Gamma $ acting freely and properly 
discontinuously in
the covering space. This group acts according to a type of generalization of
Poincar\'e's theorem - the tessellations of $H^3$ by hyperbolic
polyhedra \cite{best}. The polyhedra are pasted together, filling
up the entire $H^3$, without leaving any empty space. The motions of the
polyhedra, are performed by the discrete group of isometries $\Gamma $.
For a review of topology in connection with cosmology, see \cite{LR}
and references therein.
Among the first applications of the topology considerations, was an attempt 
to explain multiple quasar images \cite
{fagundes}. Constraints due to the homogeneity of the CMBR set 
a lower limit for the size of the fundamental cell today to $\sim 3000Mpc$ 
\cite{sok}, \cite{as} and \cite{ast}. The results apply only to
compactifications of flat space. There are arguments, however, which allow for a
compact hyperbolic manifold as the space section \cite{css}. 
The effect of the topology induces the formation
of circles in the sky, which, in principle could be measured in the CMBR 
\cite{css2}.

A very attractive argument in favor of compact hyperbolic manifolds is
connected with pre-inflationary homogenization through chaotic mixing, as
was discussed by Cornish et al. \cite{css3}. They suggested
that inflation occurred near the Planck era, after the chaotic
homogenization, so that the two processes together are better suited to
explain the large scale structure of the Universe observed today. The 
motivation for this work is based on this argument.

Due to the particular interest in the small volume hyperbolic Universe, 
we make a numerical calculation of the vacuum expectation value
of the energy of a conformally coupled massive scalar field (e.g., inflaton) 
in a static
space-time $R\times {\cal M}$, where the spatial section is the compact hyperbolic 3-manifold of
the smallest volume known, $V=0.942707...R_{\mbox{{\tiny CURV}}}^3$, 
called the Weeks 
manifold \cite{weeks}, ${\cal M}$, where $R_{\mbox{{\tiny CURV}}}$ is the 
radius of curvature of the Universe. The quantum field 
theoretical effects were strongest in the era under
consideration by Cornish et al. \cite{css3} than in any other. Previously, 
the Casimir energy for static multiply
connected flat space-times, was obtained in DeWitt et al. \cite{dewish}. We 
calculate a generalization of their result for a particular multiply
connected hyperbolic space section. We use the point splitting technique
which is very useful in determining the ultraviolet
behavior in curved space, since it involves the evaluation of field
quantities infinitesimally displaced. The obtained propagator 
is exact and possesses information about the global properties of the 
manifold in the sense that the infrared modes are taken into account, 
for Lie groups, as well as some homogeneous
space-times, such as the static $R\times S^3$ and $R\times H^3$ \cite
{camporesi}. When the space-time is multiply connected, the propagator is
obtained as the usual sum over paths: all the nontrivial geodesics connecting 
the two points are taken into account, which is the technique
used in \cite{dewish}, and the one used in this work.

We find a static hyperbolic solution for the EQ, whose spatial 
section is the Weeks manifold in section II.   
In section III, we write the expression for the vacuum expectation value of the 
energy momentum tensor. We obtain the numerical values 
of the Casimir energy in the multiply 
connected static space time of section II, in section IV. Our conclusions are 
presented in section V. (We use natural units, 
$G=c=\hbar=1$, except in section II.)
\section{The Static Universe $R\times {\cal M}$ \label{s1}}

It is well known that hyperbolic geometry can be obtained via ``Minkowski''
space, together with an additional constraint 
\begin{eqnarray}
\ &&dl^2=dx^2+dy^2+dz^2-dw^2,  \nonumber \\
\ &&\left( x-x^{\prime }\right) ^2+\left( y-y^{\prime }\right) ^2+\left(
z-z^{\prime }\right) ^2-\left( w-w^{\prime }\right) ^2=-R_{\mbox{{\tiny CURV}}}^2.  \label{cond}
\end{eqnarray}
It can be easily seen that the isometry group is the proper, 
orthocronous Lorentz
group, also called $SO(1,3),$ which is isomorphic to $
PSL(2,C)=SL(2,C)/\{\pm I\}$ \cite{dfn}. This space is homogeneous in the sense that
every point in it can be reached from any other by the action of an element of the
isometry group. Using the constraint in the line element of Eq. (\ref{cond}) 
we obtain 
\begin{eqnarray}
dl^2 &=&dx^2+dy^2+dz^2  \nonumber \\
&& -\frac{\left( \left( x-x^{\prime }\right) dx+\left( y-y^{\prime
}\right) dy+\left( z-z^{\prime }\right) dz\right) ^2}{\left( x-x^{\prime
}\right) ^2+\left( y-y^{\prime }\right) ^2+\left( z-z^{\prime }\right) ^2+
R_{\mbox{\tiny CURV}}^2%
},  \nonumber \\
dl^2 &=&g(x,x^{\prime })_{\mu \nu }dx^\mu dx^\nu ,  \label{elxp}
\end{eqnarray}
where we interchangeably write $\left( x^1,x^2,x^2\right)
\longleftrightarrow \left( x,y,z\right) $. Both connections, $\nabla _x$ and 
$\nabla _{x^{\prime }}$, compatible with the metric of Eq. (\ref{elxp}), can be
defined as
\begin{eqnarray}
&&\nabla _\mu g(x,x^{\prime })_{\alpha \beta }=\nonumber\\
&&\frac \partial {\partial
x^\mu }g(x,x^{\prime })_{\alpha \beta }-\Gamma _{\alpha \mu }^\nu
g(x,x^{\prime })_{\nu \beta }-\Gamma _{\beta \mu }^\nu g(x,x^{\prime
})_{\alpha \nu }\equiv 0,  \label{cn1} \\
&&\nabla _{\mu ^{\prime }}g(x,x^{\prime })_{\alpha \beta }=\nonumber\\
&&\frac \partial
{\partial x^{\prime \mu }}g(x,x^{\prime })_{\alpha \beta }-\Gamma _{\alpha
\mu }^{\prime \nu }g(x,x^{\prime })_{\nu \beta }-\Gamma _{\beta \mu
}^{\prime \nu }g(x,x^{\prime })_{\alpha \nu }\equiv 0,  \label{cn2}
\end{eqnarray}
where $\Gamma ^{\prime }$ means that all derivatives are taken with respect to $%
x^{\prime }.$ Spherical coordinates and the substitution 
\begin{equation}
r=\sqrt{\left( x-x^{\prime }\right) ^2+\left( y-y^{\prime }\right) ^2+\left(
z-z^{\prime }\right) ^2}\label{vlr} 
\end{equation}
in Eq. (\ref{elxp}) yield the popular Robertson-Walker line element, written in 
the Lobatchevsky form 
\begin{eqnarray}
&&ds^2 =-dt^2+R_{\mbox{\tiny CURV}}^2\left[ d\chi ^2+\sinh ^2\chi \left( d\theta ^2+\sin ^2\theta d\phi
^2\right)\right] ,  \label{ell} \\
&&\sinh ^2\chi=\frac{r^2}{R_{\mbox{\tiny CURV}}^2}=\frac{\left( x-x^{\prime }\right) ^2+\left(
y-y^{\prime }\right) ^2+\left( z-z^{\prime }\right) ^2}{R_{\mbox{\tiny CURV}}^2}.  \nonumber
\end{eqnarray}

As is well known, the EQ for the homogeneous and isotropic space sections in Eq. (%
\ref{ell}) with $R_{\mbox{\tiny CURV}}=R_{\mbox{\tiny CURV}}(t)$, reduces to the Friedmann equations 
\begin{eqnarray*}
&&\left( \frac{\dot R_{\mbox{\tiny CURV}}}{R_{\mbox{\tiny CURV}}}\right) ^2-\frac{1}
{R_{\mbox{\tiny CURV}}^2} =\frac{8\pi G}3\rho
+\frac \Lambda 3, \\
&&2\left( \frac{\ddot R_{\mbox{\tiny CURV}}}{R_{\mbox{\tiny CURV}}}\right) 
+\left( \frac{\dot R_{\mbox{\tiny CURV}}}{R_{\mbox{\tiny CURV}}}\right) ^2-\frac{1}
{R_{\mbox{\tiny CURV}}^2} =-8\pi Gp+\Lambda ,
\end{eqnarray*}
where the right hand side is the classical energy momentum source for the
geometry $T^{\mu \nu }=(\rho+p) u^\mu u^\nu +pg^{\mu \nu }$, and a 
cosmological constant
$\Lambda g^{\mu \nu }$. 

During the stages just after the Planck era, it
is likely that the Universe was radiation-dominated, so that we have the
equation of state $\rho /3=p$. Thus the energy density
scaled as $\rho =\rho_0 (R_{\mbox{\tiny CURV}0}/R_{\mbox{\tiny CURV}}(t))^4$. 
We define $8\pi G\rho_0R_{\mbox{\tiny CURV}}^4/3=C$.
By imposing that $\dot{R}_{\mbox{\tiny CURV}}=\ddot{R}_{\mbox{\tiny CURV}}=0$ for a static Universe, we obtain from 
the above 
\begin{eqnarray*}
&&\frac{C}{R_{\mbox{\tiny CURV}}^2}+1 +\frac{\Lambda}{3}R_{\mbox{\tiny CURV}}^2 =0, \\
&&-C+\frac \Lambda 3R_{\mbox{\tiny CURV}}^4=0,
\end{eqnarray*}
which have a solution  
\begin{eqnarray}
&&R_{\mbox{\tiny CURV}} =\sqrt{\frac 3{2|\Lambda |}},\nonumber \\
&&\rho =\frac \Lambda {8\pi G},\nonumber \\
&&ds^2 =-dt^2+R_{\mbox{\tiny CURV}}^2\left[ d\chi ^2+\sinh ^2\chi \left( d\theta ^2+\sin ^2\theta d\phi
^2\right)\right]\label{elpc} ,
\end{eqnarray}
where the cosmological constant is negative. 

\begin{figure}[tbph]
\centerline{\ \vspace{0.5cm} \epsfxsize=3.5cm  \epsffile{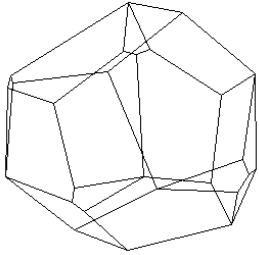}}
\caption{Fundamental region with $18$ faces in Klein coordinates, for the 
Weeks' manifold, the smallest volume manifold known.}
\label{1}
\end{figure}

According to quantum cosmology, a smaller Universe has a greater
probability of being spontaneously created. Also, the chaotic mixing
becomes more significant in this case \cite{css3}. The smallest known
compact 3-manifold ${\cal M}$, with volume $V=0.942707...R_{\mbox{\tiny CURV}}^3$ was discovered
by Weeks \cite{weeks}. It is a multiply connected manifold with universal
covering $H^3$, ${\cal M}=H^3/\Gamma $, with group $\Gamma \subset SO(1,3)$,
a discrete finite subgroup with no fixed point. Group $\Gamma $ is
isomorphic to $ \pi _1({\cal M})$, the first homotopy group,
also called the fundamental group of ${\cal M}$. $\pi _1({\cal M})$ 
is the group of
nontrivial loops composed of the maps of the manifold to the sphere ${\cal M}%
\rightarrow S^1$ \cite{dfn2}. For this smallest volume manifold, 
the $18$ $SO(1,3)$ matrices $g_i$, which generate $\Gamma (0.942707...R_{\mbox{\tiny CURV}}^3)$, were
obtained with the computer program SnapPea \cite{sp}. The fundamental domain 
is shown in FIG. \ref{1}. We note that 
the isometries preserve the form of the spatial part of the metric in 
Eq. (\ref{ell}), so that the Friedmann equations
remain unaltered since they depend only on time $t$, and the solution of EQ is the 
same as in Eq. (\ref{elpc}).\\ \\

\section{The Vacuum Expectation Energy in The Covering Space}

We now wish to evaluate the vacuum expectation value for the energy density for 
the case of a Universe consisting of a classical radiation fluid, 
a cosmological constant, and a non interacting quantum scalar field. 
The solution of EQ is given in Eq. (\ref{elpc}), where the spatial part is written 
as Eq. (\ref{elxp}) and the topology is 
$R\times H^3/\Gamma (0.942797...R_{\mbox{\tiny CURV}}^3)$. We use the point splitting method in the 
Universal covering space $R\times H^3$, for which the propagator is exact.   
The point splitting method was constructed to obtain the renormalized (finite) expectation
values of the quantum mechanical operators. It is based on the Schwinger formalism 
\cite{Schwinger} and was developed in the context of curved space by DeWitt 
\cite{dWitt}. Further details are contained in the articles of
Christensen \cite{chris1}, \cite{chris2}. For a review, see \cite{GMM}.

Metric variations in the scalar action $S$ with conformal coupling $%
\xi =1/6,$ 
\[
S=-\frac 12\int \sqrt{g}(\phi _{,\rho }\phi ^{,\rho }+\xi R\phi ^2+m^2\phi
^2)d^4x, 
\]
give the classical energy momentum tensor 
\begin{eqnarray}
T_{\mu \nu } &=&\frac 23\phi _{,\mu }\phi _\nu -\frac 16\phi _\rho \phi
^{,\rho }g_{\mu \nu }-\frac 13\phi \phi _{;\mu \nu }  \nonumber
\\
&&+\frac 13g_{\mu \nu }\phi \Box \phi +\frac 16\phi ^2G_{\mu \nu }-\frac
12m^2g_{\mu \nu },  \label{tmunu}
\end{eqnarray}
where $G_{\mu \nu }$ is the Einstein tensor. As expected for massless
fields, where $m=0$, it can be verified that the trace of the above tensor is
identically zero. Variations with respect to $\phi $ result in the curved
space generalization of the Klein-Gordon equation, 
\begin{equation}
\Box \phi -\frac R6\phi -m^2\phi =0.  \label{ekg}
\end{equation}

The renormalized energy momentum tensor
involves field products at the same space-time point. Thus the idea is to
calculate the products at separated points, $x$ and $x^\prime$, taking 
the limit at the end $x\rightarrow x^\prime$: 
\begin{equation}
\langle 0|T_{\mu \nu }|0\rangle \sim \lim_{x\rightarrow x^{\prime }}\nabla
_\mu \nabla _{\nu ^{\prime }}\frac 12\langle 0| (\phi (x)\phi (x^{\prime
})+\phi (x^{\prime })\phi (x))|0\rangle ,  \label{ordg1}
\end{equation}
where the covariant derivatives are defined in Eqs. (\ref{cn1}) and 
(\ref{cn2}).

We introduce the causal Green function 
\[
G(x,x^{\prime })=i\langle 0|T\phi (x)\phi (x^{\prime })|0\rangle , 
\]
where $T$ is the time ordering operator. Taking the real and imaginary 
parts of the Feynman propagator, 
\begin{equation}
G(x,x^{\prime })=G_s(x,x^{\prime })+\frac i2G^{(1)}(x,x^{\prime }),
\label{dfg}
\end{equation}
results in the Hadamard function 
\[
G^{(1)}(x,x^{\prime })=\langle 0|\{\phi (x),\phi (x^{\prime })\}|0\rangle , 
\]
which is the expectation value of the anti-commutator 
$\{\phi (x),\phi (x^{\prime })\}=\phi(x)\phi(x^\prime)+\phi(x^\prime)\phi(x)$, which
appears in Eq. (\ref{ordg1}). By the above reasoning, taking into account 
Eq. (\ref{ekg}) and the field products in the classical equation (\ref{tmunu}), 
the vacuum expectation value is obtained
\begin{eqnarray}
&&\langle 0|T_{\mu \nu }|0\rangle =\nonumber\\&&\lim_{x\rightarrow x^{\prime }}\left[
\frac 16\left( \nabla _\mu \nabla _{\nu ^{\prime }}+\nabla _{\mu ^{\prime
}}\nabla _\nu \right)
-\frac 1{12}g(x,x)_{\mu \nu }\nabla _\rho \nabla ^{\rho
^{\prime }}\right.\nonumber\\
&&\left.-\frac 1{12}\left( \nabla _\mu \nabla _\nu +\nabla _{\mu ^{\prime
}}\nabla _{\nu ^{\prime }}\right)+
\frac 1{48}g(x)_{\mu \nu }\left( \Box +\Box ^{\prime }\right)\right.\nonumber\\
&&\left. +\frac
1{12}\left( R(x)_{\mu \nu }
-\frac 14R(x)g(x)
_{\mu \nu }\right) -\frac 18m^2g(x)_{\mu \nu
}\right] \nonumber\\&&G^{(1)}(x,x^{\prime }),  \label{Tmn} \\
&&\langle 0|T_{\mu \nu }|0\rangle =\lim_{x\rightarrow x^{\prime }}\left[
T(x,x^{\prime })_{\mu \nu }\right] ,  \label{Tnl}
\end{eqnarray}
where the metric used to abtain the curvature tensor $R(x)_{\mu\nu}$, 
$g(x)_{\mu\nu}=g(x,x^\prime=0)_{\mu\nu}$, is given in  (\ref{elxp}), 
and the covariant derivatives are given in (\ref{cn1}) and (\ref{cn2}).
\section{The Feynman Propagator and the Casimir Energy in 
$R\times {\cal M}$}

The Feynman propagator involves a summation over all possible classical 
paths. When the space is multiply connected, it is necessary to sum over 
all nontrivial geodesics connecting two points. 
Green functions are solutions of Eq. (\ref{ekg}) with the
boundary condition 
\begin{eqnarray*}
&&F(x,x^{\prime })G(x,x^{\prime }) =\delta (x-x^{\prime }), \\
&&F(x,x^{\prime }) =F(x)/\sqrt{-g}\delta (x-x^{\prime }),
\end{eqnarray*}
where $F(x)=\Box -R/6-m^2.$ 

We introduced an auxiliary evolution
parameter $s$ and a complete orthonormal set of states $|x\rangle$, such
that 
\begin{eqnarray*}
&&G(x,x^{\prime }) =\langle x|\hat G|x^{\prime }\rangle , \\
&&F(x,x^{\prime }) =\langle x|\hat F|x^{\prime }\rangle , \\
&&\hat F\hat G =\hat I.
\end{eqnarray*}
This last equation implies that $\hat G=(\hat
F-i0)^{-1}$, so that the causal Green function becomes
\begin{equation}
G(x,x^{\prime })=i\int_0^\infty ds\langle x|\exp (-is\hat F)|x^{\prime
}\rangle   \label{int}
\end{equation}
and the matrix element $\langle x|\exp (-is\hat F)|x^{\prime }\rangle
=\langle x(s)|x^{\prime }(0)\rangle $ satisfies a Schr\"odinger type
equation, 
\[
i\frac \partial {\partial s}\langle x(s)|x^{\prime }(0)\rangle =\left( \Box
-R/6-m^2\right) \langle x(s)|x^{\prime }(0)\rangle .
\]

Dowker and Critchley \cite{dw1} obtained the Green function for the static 
homogeneous space-time with spherical
space sections $(S^3)$, using the above technique. Using a similar 
procedure, the result for a static hyperbolic space
section is obtained.
Assuming that $\langle x(s)|x^{\prime }(0)\rangle $ depends only on the
geodesic distance $\chi $, given in Eq. (\ref{ell}), the above
equation is easily solved. 
By substituting the solution 
$\langle x(s)|x^{\prime }(0)\rangle $ for the integrand in Eq. (\ref{int}),
\begin{equation}
G(x,x^{\prime })=-\frac m{8\pi }\frac \chi {\sinh \chi }\frac{%
H_1^{(2)}\left( m\sqrt{\left( t-t^{\prime }\right) ^2-
R_{\mbox{{\tiny CURV}}}^2\chi ^2}\right) }{%
\sqrt{\left( t-t^{\prime }\right) ^2-R_{\mbox{{\tiny CURV}}}^2\chi ^2}},  \label{ff}
\end{equation}
where $H_1^{(2)}$ is the Hankel function of the second kind of order $1$,
the causal Green function is obtained.

The Klein-Gordon equation remains unchanged under isometries, 
\[
\mbox{$\pounds _\xi$}\left[ \left( \Box -\frac R6-m^2\right) \phi \right] =
\left(\Box -\frac R6-m^2\right) \mbox{$\pounds _\xi$} \phi ,
\]
where $\mbox{$\pounds_\xi$}$ is the Lie derivative with respect to 
$\xi$, is the Killing vector that generates the isometry,
so that summations in the Green functions over the discrete elements of the 
group $\Gamma(0.942707...R_{\mbox{{\tiny CURV}}}^3)$ remains well defined.  
For comparison, we have from Eq. (6) in \cite{dw1},  
\begin{eqnarray*}
&&G(x,x^{\prime },\kappa ^2)=-\frac{\kappa ^2}{8\pi a\sin (s/a)}\nonumber\\&&
\sum_{n=-\infty }^\infty \left( s+2\pi na\right) \frac{H_1^{(2)}\left(
\kappa \sqrt{(t-t^{\prime })^2-s^2}\right) }{\kappa \sqrt{(t-t^{\prime
})^2-s^2}}. 
\end{eqnarray*}
 
The infinite summation appears because the space sections $S^3$ are compact
and, therefore, there are an infinite number of geodesics connecting two points. In fact, 
by removing
the summation, leaving only the direct path and making the substitutions $%
s/a\rightarrow i\chi $, $a\rightarrow iR_{\mbox{{\tiny CURV}}},$ each formula may be derived from the 
other. The Hadamard function can be obtained from Eqs. (\ref{ff}) and (\ref{dfg}), 
\begin{equation}
G^{(1)}(x,x^{\prime }) =\frac m{2\pi ^2}\frac \chi {\sinh \chi }\frac{%
K_1\left( m\sqrt{-\left( t-t^{\prime }\right) ^2+R_{\mbox{{\tiny CURV}}}^2\chi ^2}\right) }{\sqrt{%
-\left( t-t^{\prime }\right) ^2+R_{\mbox{{\tiny CURV}}}^2\chi ^2}},  \label{Hdrmf}
\end{equation}
where $K_1$ is the modified Bessel function of the second kind.

Substituting Eq. (\ref{Hdrmf}) and the covariant derivatives (\ref{cn1}) and(\ref
{cn2}) in Eq. (\ref{Tmn}), we obtain  
$
T(x,x^{\prime })_{\mu \nu }
$
in Eq. (\ref{Tnl}). In ${\cal M}=R\times H^3/\Gamma (0.942707...R_{\mbox{{\tiny CURV}}}^3)$,
the summation over the infinite geodesics connecting the two points $x$ and $%
x^{\prime }$ is performed by the action of the generators $g_i$ of the group $%
\Gamma (0.942707...R_{\mbox{{\tiny CURV}}}^3)$ and their products 
on the points $x^{\prime }$, (for example, in (\ref{Tnl})):  
\begin{equation}
\left\langle 0\left| T(x,0.942707...
R_{\mbox{{\tiny CURV}}}^3)_{\mu \nu }\right| 0\right\rangle\
=\lim_{x\rightarrow x^{\prime }} 
\sum_{i}T\left(
x,\Gamma_ix^{\prime }\right) _{\mu \nu }.  \label{resl}
\end{equation}

We evaluated Eq. (\ref{resl}) numerically for the compact 
space-time ${\cal M}$.
In the summation, the direct path gives a divergent contribution.
It can be shown that avoiding the direct path is equivalent to a 
renormalization of the cosmological constant \cite{GMM}. We shall use the 
same type of renormalization.
In Eq. (\ref{resl}), we summed over the generators and their products up to 
three factors (see below) and assured
that no transformed point in the covering space $\Gamma_ix^\prime$ is 
summed more than once. We also checked that relation
(\ref{cond}) was verified for each transformed point, $\Gamma_ix^\prime$. 

\begin{figure}[tbph]
\centerline{\ \epsfxsize=8cm \vspace{0.5cm} \epsffile{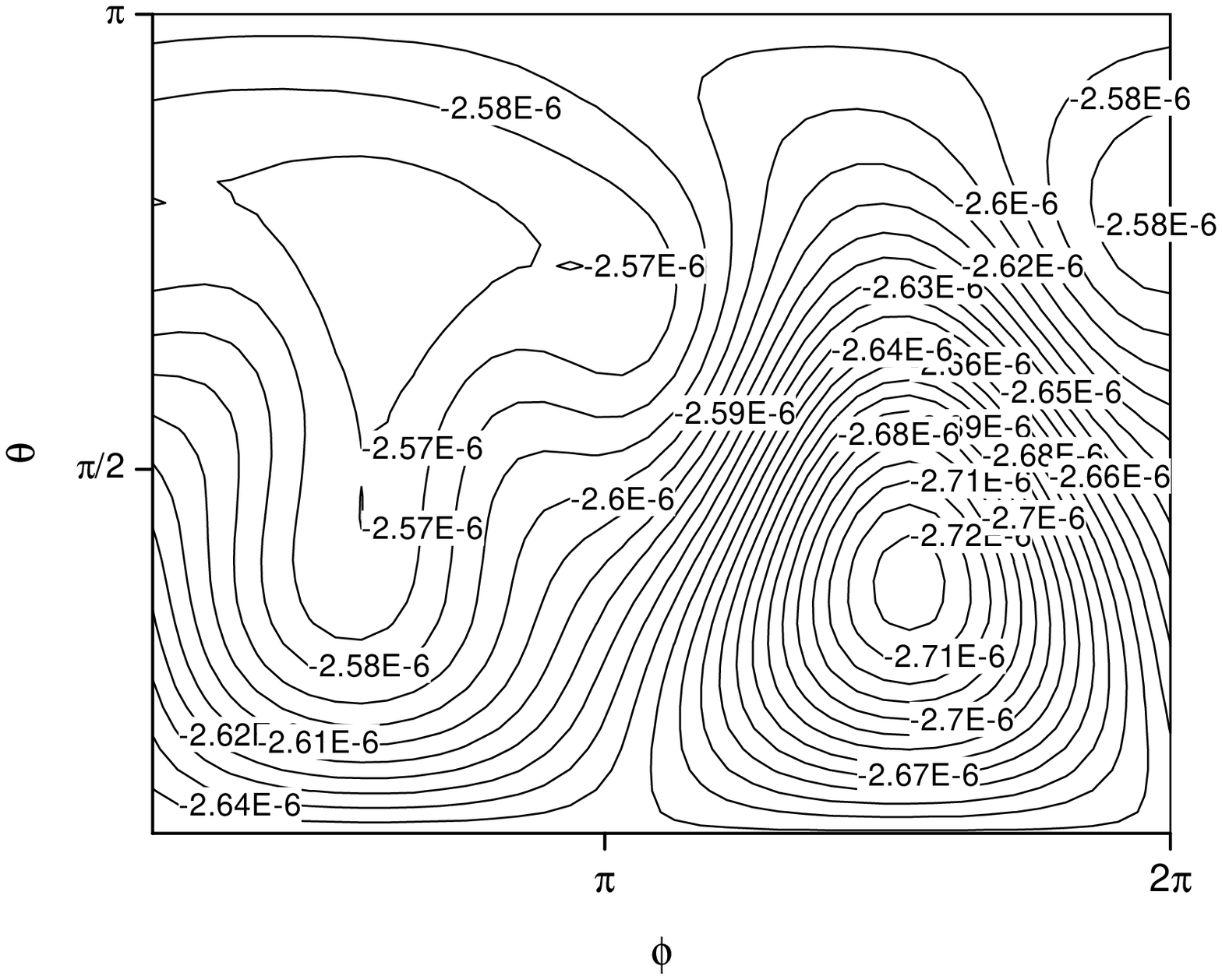}}
\caption{Renormalized value of the vacuum $00$ component of a conformally 
coupled massive scalar field $\phi $ $(E=T_{\mu \nu }u^\mu u^\nu =T_{00})$ 
with a mass $m=0.5$
in the static $R\times {\cal M}$ space-time with $R_{\mbox{{\tiny CURV}}}=10$.  
The $T_{00}$ is on a sphere of radius $r=0.6$, inside the fundamental polyhedron, 
FIG. \ref{1}, where $r$ is given in Eq. (\ref{vlr}).}
\label{2}   
\end{figure}

We obtained the result shown in FIG. \ref{2}, for a scalar field with 
mass $m=0.5$,  $R_{\mbox{{\tiny CURV}}}=10$. The values of the vacuum 
energy, E, in FIG. \ref{2}, seen by an 
observer with a four velocity $u^\mu=(1,0,0,0)$ of \[E= 
\left\langle 0\left| T(x,0.942707...R_{\mbox{{\tiny CURV}}}^3)_{\mu \nu }\right| 0\right\rangle 
u^\mu u^\nu\]  at 
each point $x$ on the surface of a sphere inside the fundamental polyhedron 
in FIG. \ref{1}. The radius of the sphere is chosen to be $r=0.6$, 
where $r$ is given in Eq.
(\ref{vlr}). $\theta$ and $\phi$ correspond to the latitude and longitude, 
so that the lines $\theta=0$ and $\theta=\pi$ correspond to the 
south and north poles, respectively. It is clear from FIG. \ref{2} 
that there are 
spontaneous vacuum excitations of low multipolar components. 

The infinite summation in Eq. (\ref{resl}) occurs 
because the space-time $R\times {\cal M}$ is static, so that there has  
been enough time for the quantum interaction of the scalar field with the 
geometry to travel through space. Since the Universe is expanding, 
we know that the summation ceases to be strictly physically valid. The presence 
of the mass term, however, naturally introduces a cutoff.

In Eq. (\ref{resl}), we summed over the generators and their products. We 
found that products of four generators or more, contributed less than 
$10^{-14}$ to the values in FIG. \ref{2} (i.e. $\sim 10^{-8}$ of the 
plotted values of $T_{00}$).
\section{Conclusions}
We explicitly evaluated the distribution of the vacuum (Casimir) energy of 
a conformally coupled massive scalar field for a specific interesting 
situation: the smallest static Universe of negative curvature known, whose spatial 
sections are the Weeks manifold. As a specific example, we chose $m=0.5$ for 
the mass of the scalar field, $R_{\mbox{{\tiny CURV}}}=10$ for the radius of 
curvature of the Universe and values of $T_{00}$ on a sphere of radius 
$r=0.5$ inside the fundamental polyhedron FIG. \ref{1}, are shown in FIG. \ref{2}. It can 
be seen in FIG. \ref{2} that there is a maximum of 
$T_{00}\sim 2.6\times 10^{-6}$ at $\theta\sim3\pi/4$, $\phi\sim\pi/2$ and a 
minimum of $T_{00}\sim 2.7\times 10^{-6}$ at $\theta\sim \pi/4$, 
$\phi\sim 3\pi/2$. It is to be noted that this type of topological effect 
depends on the mass: for smaller values of $m$ the maximum values are bigger 
and the minimum values are smaller.
\section*{Acknowledgments} 
D.M. would like to thank the Brazilian agency FAPESP for financial 
support. H. V. F. thanks the Brazilian agency CNPq for partial support. 
R.O. thanks FAPESP and CNPq for partial support.

\end{document}